\documentclass[preprint2]{aastex}

\newcommand{\lae}{\mathrel{<\kern-1.0em\lower0.9ex\hbox{$\sim$}}}
\newcommand{\gae}{\mathrel{>\kern-1.0em\lower0.9ex\hbox{$\sim$}}}
 
\shorttitle{The 20-40 keV AGN Luminosity Function}
\shortauthors{Beckmann et al.}

\begin{document}

\title{The Hard X-ray 20-40 keV AGN Luminosity Function}


\author{V. Beckmann\altaffilmark{1}}
\affil{NASA Goddard Space Flight Center, Exploration of the Universe Division, Code 661, Greenbelt, MD 20771, USA}
\email{beckmann@milkyway.gsfc.nasa.gov}
\author{S. Soldi\altaffilmark{2,3}, C. R. Shrader\altaffilmark{4,5}, N. Gehrels\altaffilmark{4}, and N. Produit\altaffilmark{2}}


\altaffiltext{1}{also with the Joint Center for Astrophysics, Department of Physics, University of Maryland, Baltimore County, MD 21250, USA}
\altaffiltext{2}{INTEGRAL Science Data Centre, Chemin d' \'Ecogia 16, 1290
 Versoix, Switzerland}
\altaffiltext{3}{also with Observatoire de Gen\`eve, 51 Ch. des Maillettes, 1290 Sauverny, Switzerland}
\altaffiltext{4}{NASA Goddard Space Flight Center, Exploration of the Universe Division, Code 661, Greenbelt, MD 20771, USA}
\altaffiltext{5}{also with Universities Space Research Association, 10211 Wincopin Circle, Columbia, MD 21044, USA}

\begin{abstract}
We have compiled a complete
extragalactic sample based on $\sim25,000 \rm \, deg^2$ to a limiting flux of $3 \times 10^{-11} \rm \, ergs \, cm^{-2} \, s^{-1}$ ($\sim7,000 \rm \, deg^2$ to a flux limit of $10^{-11} \rm \, ergs \, cm^{-2} \, s^{-1}$) in the 20 -- 40 keV band with {\it INTEGRAL}. We have constructed a detailed exposure map to compensate for effects of non-uniform exposure. The flux-number relation is best described by a power-law with a slope of $\alpha = 1.66 \pm 0.11$. The integration of the cumulative flux per unit area leads to $f_{20 - 40 \rm \, keV} = 2.6 \times 10^{-10} \rm \, ergs \, cm^{-2} \, s^{-1} \, sr^{-1}$, which is about 1\% of the known $20 - 40 \rm \, keV$ X-ray background.
We present the first luminosity function of AGN in the 20--40 keV energy range, based on 38 extragalactic objects detected by the imager IBIS/ISGRI on-board {\it INTEGRAL}. The luminosity function shows a smoothly connected two power-law form, with an index of $\gamma_1 = 0.8$ below, and $\gamma_2 = 2.1$ above the turn-over luminosity of $L_* = 2.4 \times 10^{43} \rm \, ergs \, s^{-1}$. The emissivity of all {\it INTEGRAL} AGNs per unit volume is $W_ {20 - 40 \rm \, keV}(> 10^{41} \rm \, ergs \, s^{-1}) = 2.8 \times 10^{38} \rm \, ergs \, s^{-1} \, h_{70}^3 \, Mpc^{-3}$. These results are consistent with those derived in the $2 - 20 \rm \, keV$ energy band and do not show a significant contribution by Compton-thick objects. Because the sample used in this study is truly local (${\bar z} = 0.022$), only limited conclusions can be drawn for the evolution of AGNs in this energy band. 
\end{abstract}




\section{Introduction}

The Galactic X-ray sky is dominated by accreting binary systems, while the extragalactic sky shows mainly active galactic nuclei (AGN) and clusters of galaxies. Studying the population of sources in X-ray bands has been a challenge ever since the first observations by rocket borne X-ray detectors (Giacconi et al. 1962). At soft X-rays (0.1 -- 2.4 keV) deep exposures by {\it ROSAT} have revealed an extragalactic population of mainly broad line AGNs, such as type Seyfert 1 and quasars \citep{Hasinger98,ROSATdeep}. 
In the 2 - 10 keV range surveys 
have been carried out with {\it ASCA} (e.g. Ueda et al. 2001), {\it XMM-Newton} (e.g. Hasinger 2004), and {\it Chandra} (e.g. Brandt et al. 2001) and have shown that the dominant extragalactic sources are more strongly absorbed than those within the {\it ROSAT} energy band.
For a summary on the deep X-ray surveys below 10 keV see Brandt \& Hasinger (2005). At higher energies the data become more scarce. Between a few keV and $\sim 1 \rm \, MeV$, no all-sky survey using imaging instruments has been performed to date. The {\it Rossi X-ray Timing Explorer} ({\it RXTE}) sky survey in the $3 - 20 \rm \, keV$ energy band revealed about 100 AGNs, showing an even higher fraction of absorbed ($N_H > 10^{22} \rm \, cm^{-2}$) sources of about 60\% \cite{RXTENGC}. 
The {\it International Gamma-Ray Astrophysics Laboratory} ({\it INTEGRAL}; Winkler et al. 2003) offers an unprecedented $>20 \rm \, keV$ collecting area and state-of-the-art detector electronics and background rejection capabilities. Notably, the imager IBIS with an operating range from $20 - 1000 \rm \, keV$ and a fully-coded field of view of $10^\circ \times 10^\circ$ enables us now to study a large portion of the sky. A first catalog of AGNs showed a similar fraction of absorbed objects as the {\it RXTE} survey \cite{intagn}. 
The Burst Alert Telescope (BAT) of the {\it Swift} mission \cite{Swift} operates in the 15 -- 200 keV band and uses a detector similar to IBIS/ISGRI, but provides a field of view about twice the size. The BAT data of the first three months of the mission provided a high galactic latitute survey, including 44 AGNs \cite{BATsurvey}. Within this sample a weak anti-correlation of luminosity versus intrinsic absorption was found as previously found in the $2-10 \rm \, keV$ band \citep{Ueda03,LaFranca}, revealing that most of the objects with luminosities $L_X > 3 \times 10^{43} \rm \, ergs \, s^{-1}$ show no intrinsic absorption. Markwardt et al. (2005) also pointed out that this luminosity corresponds to the break in the luminosity function. 

Related to the compilation of AGN surveys in the hard X-rays is the question of what sources form the cosmic X-ray background (CXB). While the CXB below 20 keV has been the focus of many studies, the most reliable measurement in the 10 - 500 keV has been provided by the {\it High Energy Astronomical Observatory} ({\it HEAO 1}), launched in 1977 (Marshall et al. 1980). The most precise measurement provided by the UCSD/MIT Hard X-ray and Gamma-Ray instrument ({\it HEAO 1} A-4) shows that the CXB peaks at an energy of about $30 \rm \, keV$ (Marshall et al. 1980, Gruber et al. 1999). The isotropic nature of the X-ray background points to an extragalactic origin, and as the brightest persistent sources are AGNs, it was suggested early on that those objects are the main source of the CXB (e.g. Setti \& Woltjer 1989). In the soft X-rays this concept has been proven to be correct through the observations of the {\it ROSAT} deep X-ray surveys, which showed that $90 \%$ of the $0.5 - 2.0 \rm \, keV$ CXB can be resolved into AGNs (Schmidt et al. 1998). At higher energies ($2 - 10 \rm \, keV$), {\it ASCA} and {\it Chandra} surveys measured the hard X-ray luminosity function (XLF) of AGNs and its cosmological evolution. These studies show that in this energy range the CXB can be explained by AGNs, but with a higher fraction of absorbed ($N_H > 10^{22} \rm \, cm^{-2}$) objects than in the soft X-rays (e.g. Ueda et al. 2003). 
A study based on the {\it RXTE} survey by Sazonov \& Revnivtsev (2004) derived the local hard X-ray luminosity function of AGNs in the 3--20 keV band. They showed that the summed emissivity of AGNs in this energy range is smaller than the total X-ray volume emissivity in the local Universe, and suggested that a comparable X-ray flux may be produced together by lower luminosity AGNs, non-active galaxies and clusters of galaxies. Using the {\it HEAO 1}-A2 AGNs, Shinozaki et al. (2006), however, obtained a local AGN emissivity which is about twice larger than the value of Sazonov \& Revnivtsev (2004) but consistent with the estimates by Miyaji et al. (1994) which was based on the cross-correlation of the {\it HEAO 1}-A2 map with {\it IRAS} galaxies.

With the on-going observations of the sky by {\it INTEGRAL}, a sufficient amount of data is now available to derive the AGN hard X-ray luminosity function.
In this paper we present analysis of recent observations performed by the {\it INTEGRAL} satellite, and compare the results with previous studies. In Section 2 we describe the AGN sample and in Section 3 the methods to derive the number-flux distribution of {\it INTEGRAL} AGNs are presented together with the analysis of their distribution. Section 4 shows the local luminosity function of AGNs as derived from our data, followed by a discussion of the results in Section 5. 
Throughout this work we applied a cosmology with $H_0 = 70 \rm \, km \, s^{-1} \, Mpc^{-1}$ ($h_{70} = 1$), $k = 0$ (flat Universe), $\Omega_{matter} = 0.3$, and $\Lambda_0 = 0.7$, although a $\Lambda_0 = 0$ and $q_0 = 0.5$ cosmology does not change the results significantly because of the low redshifts in our sample.

\section{The {\it INTEGRAL} AGN Sample}

Observations in the X-ray to soft gamma-ray domain have been performed by the soft gamma-ray imager (20--1000 keV) ISGRI \cite{ISGRI} on-board the {\it INTEGRAL} satellite \cite{INTEGRAL}. 

The data used here are taken from orbit revolutions 19 - 137 and revolutions 142 - 149. 
The list of sources was derived from the analysis as described in Beckmann et al. (2006a). The analysis was performed using the Offline Science Analysis (OSA)
software version 5.0 distributed by the ISDC (Courvoisier et
al. 2003a). Additional observations performed later led to further source detections within the survey area. We extracted spectra at those positions from the data following the same procedure. It is understood that most of those objects did not result in a significant detection $\ge 3 \sigma$ in the data set used here, but it ensures completeness of the sample at a significance limit of $5 \sigma$ (see Section 3).
 

The list of 73 sources is shown in Tab.~\ref{catalog}. 22 of the sources have Galactic latitudes $-10^\circ < b < +10^\circ$ (14, if we only consider the sources with significance $\ge 5\sigma$). In addition to the sample presented here, 8 new {\it INTEGRAL} sources with no identification have been detected in our survey with a significance of $\ge 5 \sigma$. These un-identified sources, most of them in the Galactic Plane, are not included in this work.   
The significances listed 
have been derived from the intensity maps produced by the OSA software. Different to Beckmann et al. (2006a) we did not use the significances as determined for the whole ISGRI energy range by the extraction software, but determined the significances based on the count rate and count rate error for ISGRI in the 20 -- 40 keV energy band only, as this is the relevant energy range for this work. 
Fluxes are determined by integrating the best-fit spectral model over the
20--40 keV bandpass. The uncertainty in the absolute flux calibration is
about 5\%.
The luminosities listed are the luminosities in this energy band, based on the measured (absorbed) flux. The absorption listed is the intrinsic absorption in units of $10^{22} \rm \, cm^{-2}$ as measured in soft X-rays below 10 keV by various missions as referenced. We also include the most important reference for the {\it INTEGRAL} data of the particular source in the last column of Table~\ref{catalog}. The extracted images and source results are available in electronic form\footnote{http://heasarc.gsfc.nasa.gov/docs/integral/inthp\_archive.html}.

In order to provide a complete list of AGNs detected by {\it INTEGRAL}, we included also those sources which are not covered by the data used for our study. Those sources are marked in Tab.~\ref{catalog} and are not used in our analysis. 



\section{Number-Flux Distribution of {\it INTEGRAL} AGNs}

\subsection{Completeness of the Sample}

In order to compute the AGN number-flux relation it is necessary to have a complete and unbiased sample. Towards this end, one must understand the characteristics of the survey, such as the sky coverage and completeness for each subset of the total sample.  
Because of the in-homogeneous nature of the survey exposure map, we applied a significance limit rather than a flux limit to define a complete sample. The task is to find a significance limit which ensures that all objects above a given flux limit have been included. To test for completeness, the 
$V_e/V_a$-statistic has been applied, where $V_e$ stands for the volume that is enclosed by the object, and $V_a$ is the accessible volume, in which the object could have been found (Avni \& Bahcall 1980).

In the case of no evolution 
$\langle V_e/V_a \rangle = 0.5$ is expected. 
This evolutionary test is applicable only to samples complete to a well-defined significance limit. 
It can therefore also be used to test the completeness of a sample. 
We performed a series of $V_e/V_a$-tests to the {\it INTEGRAL} AGN
sample, assuming completeness limits in the range of $0.5 \sigma$ up to
$16\sigma$ ISGRI 20 -- 40 keV significance. For a significance limit
below the true completeness limit of the sample one expects the
$V_e/V_a$-tests to derive a value $\langle V_e/V_a \rangle < \langle
V_e/V_a \rangle_{true}$, where $\langle V_e/V_a \rangle_{true}$ is the
true test result for a complete sample. Above the completeness limit
the $\langle V_e/V_a \rangle$ values should be distributed around
$\langle V_e/V_a \rangle_{true}$ within the statistical
uncertainties. 

\begin{figure}
\plotone{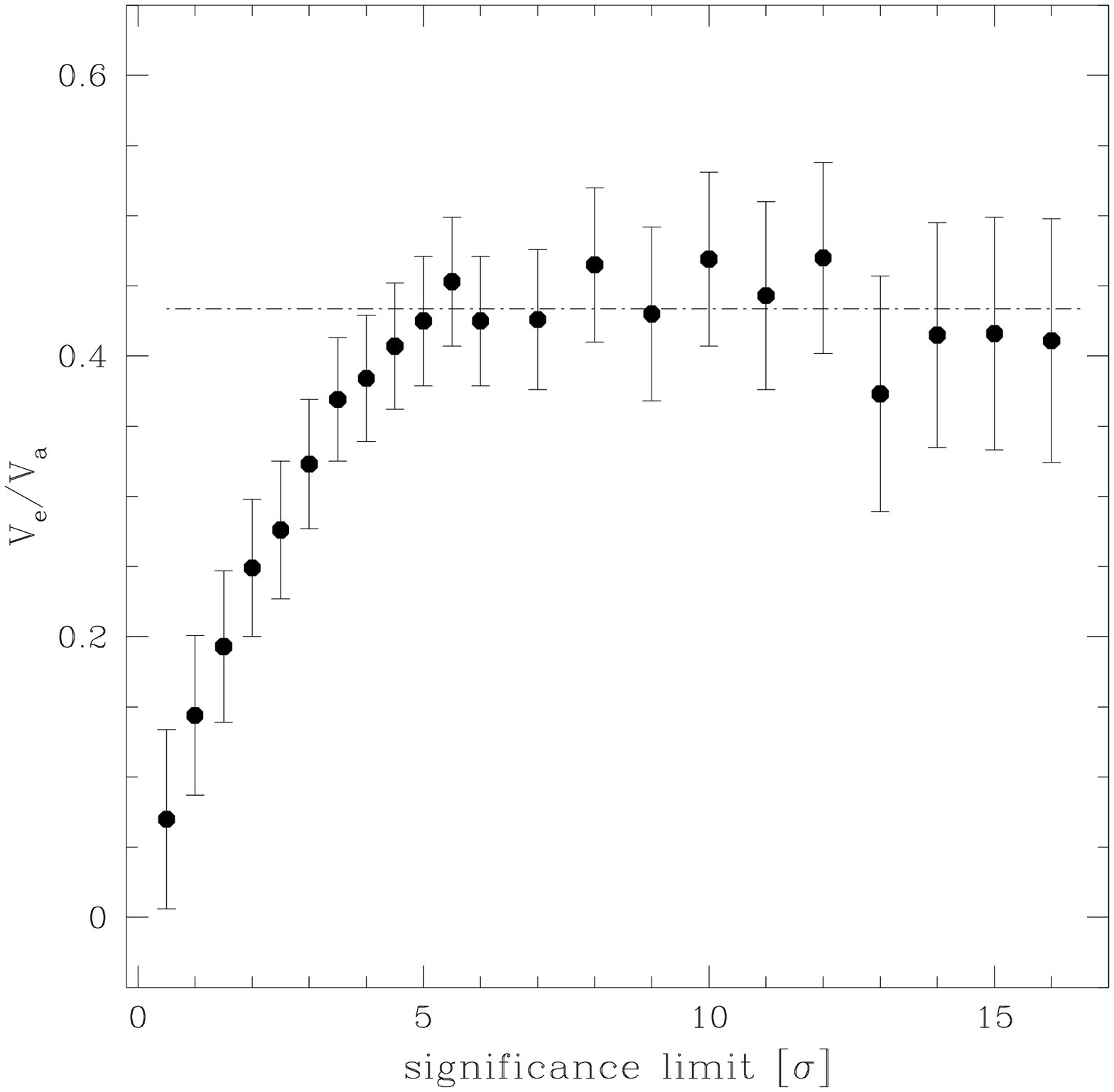}
\caption[]{Result of the $V_e/V_a$ test, assuming different completeness limits in ISGRI significance. The dot-dashed line shows the average $\langle V_e/V_a \rangle$ value for the objects with significance $\ge 5 \sigma$.}

\label{fig:vvmaxtest}
\end{figure}

The results of the tests are shown in Figure~\ref{fig:vvmaxtest}. 
It appears that the sample becomes complete at
a significance cutoff of approximately $5\sigma$, which includes 38 AGNs. The average value
is  $\langle V_e/V_a \rangle = 0.43 \pm 0.05$. This is consistent
with the expected value of 0.5 at the $1.5 \sigma$ level, suggesting no evolution and a uniform distribution in the local universe. It is unlikely that cosmological effects have an influence on the result, as the average redshift in the sample is $\bar{z} = 0.022$, with a maximum redshift of $z = 0.13$. A positive cosmological evolution would result in a slightly higher value than $0.5$. We would like to remind that we use the $\langle V_e/V_a \rangle$ test is not to determine any cosmological effects, but use it to see at what significance level it returns a stable value. 

\subsection{Deriving the Area Corrected Number-Flux Distribution}

\begin{figure}
\plotone{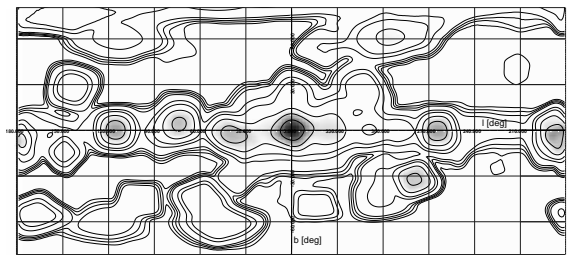}
\caption[]{Exposure map representing the data used in our analysis in Galactic coordinates. Contours indicate 2 ks, 5 ks, 10 ks, 20 ks, 100 ks, 200 ks, 500 ks, 1 Ms, 2 Ms exposure time. {\it INTEGRAL} spent most of the observing time on and near the Galactic plane, with a strong focus on the Galactic center and on areas including bright hard X-ray sources like the Cygnus region, Vela, GRS 1915+105, and the Crab. Fields at high Galactic latitude include Coma, Virgo and the region around NGC~4151.} 

\label{fig:exposuremap}
\end{figure}

\begin{figure}
\plotone{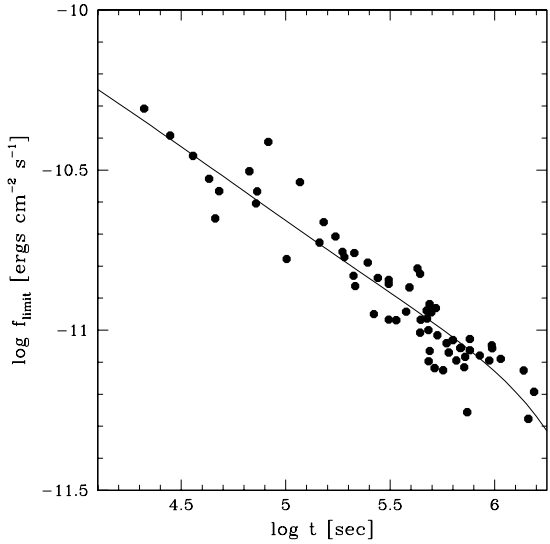}
\caption[]{Correlation of exposure time and flux limit ($5 \sigma$, $20 - 40 \rm \, keV$) for the AGNs in this study (see Sect.\ref{logNlogSslope}). The curve shows a smooth polynomial fit for flux limit versus logarithm of the exposure time.}

\label{fig:polyfit}
\end{figure}

\begin{figure}
\plotone{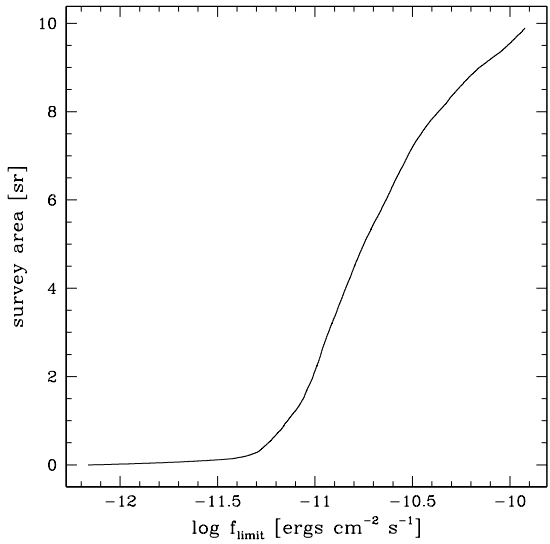}
\caption[]{The total survey area with respect to the $5\sigma$ flux limit in the 20 - 40 keV. The curve is based on all 47,868 sky elements of the survey with an exposure of at least 2 ks. For a flux limit of $f_X \ge 3 \times 10^{-11} \rm \, ergs \, cm^{-2} \, s^{-1}$ the survey covers $76 \%$ of the sky, and for $f_X \ge 10^{-11} \rm \, ergs \, cm^{-2} \, s^{-1}$ 17\%.}

\label{fig:fluxlimits}
\end{figure}

A correct representation of the number flux distribution (i.e. $\log N_{>S}$ versus $\log S$, see Beckmann et al. 2006b) for the sample presented here must account for different exposure times comprising our survey, and the resulting sensitivity variations.
We determine here the number density and thus the number of AGNs above a given flux has to be counted and divided by the sky area in which they are detectable throughout the survey. We therefore first determined the exposure time in $64,620$ sky elements of $\sim 0.63 \rm \, deg^2$ size within our survey. In each sky bin, the exposure is the sum of each individual exposure 
multiplied by the fraction of the coded field of view in this particular 
direction. The dead time and the good time intervals (GTI) are not taken into account but
the dead time is fairly constant (around 20\%) and GTI gaps are very rare in IBIS/ISGRI data. 
Figure~\ref{fig:exposuremap} shows the exposure map in Galactic coordinates for this survey.
We excluded those fields with an exposure time less than 2 ks, resulting in $47,868$ sky elements with a total coverage of $9.89 \rm \, sr$.
The flux limit for a given significance limit should be a function of the square root of the exposure time, if no systematic effects apply, but this assumption cannot be made here. The nature of coded-mask imaging leads to accumulated systematic effects at longer exposure times. In order to achieve a correlation between the exposure time and the flux limit, we therefore used an empirical approach. 
For each object we computed what we will call its $5 \sigma$ equivalent flux $f_{5 \sigma}$, based on its actual flux $f_X$ and its significance $s$: 
$f_{5 \sigma} = f_X * 5 / s$.
We found a correlation between these $f_{5 \sigma}$ values and exposure times, which has a scatter of $\lae 0.2 \rm \, dex$ (Fig.~\ref{fig:polyfit}). 
The correlation was then fitted by a smooth polynomial of third degree. This function was then used to estimate the limiting flux of each individual survey field. It must be noted that the individual limits are not important, but only the distribution of those flux limits. The total area in the survey for a given flux limit is shown in Figure~\ref{fig:fluxlimits}.    

\begin{figure}
\plotone{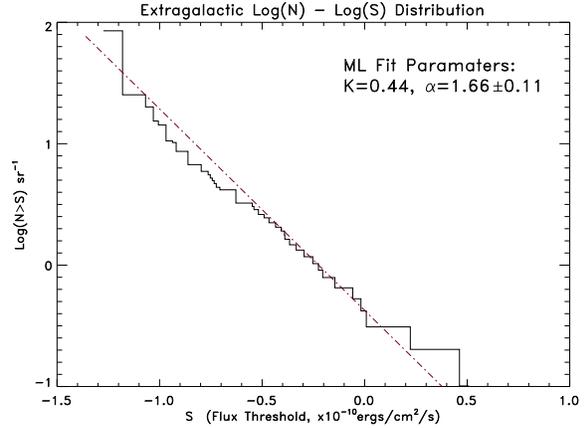}
\caption[]{Number flux distribution per steradian of {\it INTEGRAL} AGNs with a detection significance $> 5 \sigma$. Blazars have been excluded. The maximum likelihood slope as described in Section \ref{logNlogSslope} is $1.66 \pm 0.11$.}

\label{fig:logNlogS5}
\end{figure}

Based on the flux limits for all survey fields, we are now able to construct the number flux distribution for the {\it INTEGRAL} AGNs, determining for each source flux the total area in which the source is detectable with a $5\sigma$ detection significance in the $20 - 40 \rm \, keV$ energy band. The resulting correlation is shown in Figure~\ref{fig:logNlogS5}.

\subsection{The Slope of the Number-Flux Distribution}
\label{logNlogSslope}

We applied a maximum-likelihood (ML) algorithm to our
empirical number-flux distribution to obtain a 
power-law approximation
of the form $N(>S) = K \, \cdot \, S^{-\alpha}$. We note that
we are fitting the "integrated" $N(>S)$ function, as distinct from 
the "differential" number-flux function. The latter entails binning
the data, and thus some loss of information is incurred. The advantage of fitting the differential distribution is that a simple least squares procedure may be employed. However, given the modest size of our sample, the expected loss of accuracy was considered unacceptable. 

Our approach was based on the
formalism derived by Murdoch, Crawford \& Jauncey (1973), also
following the implementation of Piccinotti et al. (1982). The latter
involved modification of the basic ML  incorporated to facilitate
handling of individual source flux-measurement errors. The ML method
also involves the application of a Kolmogorov-Smirnov (K-S) 
test as part of the procedure to optimize the fit, 
as detailed in Murdoch, Crawford \& Jauncey (1973) (we note
that the KS test as applied in this context is not a measure of the
overall goodness of fit). Once 
the slope is determined, a chi-square minimization is used to 
determine the amplitude K. 

For this analysis, we used the complete sub-sample of 38 sources for which
the statistical significance of our flux determinations was 
at a level of $5 \sigma$ or greater. The dimmest source among 
this sub-sample was
$f_X = 5.6 \times 10^{-12} \rm \, ergs \, cm^{-2} \, s^{-1}$, 
and the brightest was 
$f_X = 3.2 \times 10^{-10} \rm \, ergs \, cm^{-2} \, s^{-1}$. 
We derived a ML
probability distribution, which can be approximated by a Gaussian, 
with our best fit parameters of $\alpha = 1.66 \pm 0.11$. A 
normalization of
$K=0.44 \, \rm sr^{-1} \, (10^{-10} \rm \, ergs \, cm^{-2} \, s^{-1})^{\alpha}$ 
was then obtained by performing a least-squares fit, with the 
slope fixed to the ML value. 
This calculation did not take into account possible inaccuracies 
associated with scatter in Fig.~\ref{fig:polyfit}, and thus in the true detection limit. 
The true exposure time is also affected by small variations in 
dead-time effects known to occur in the 
ISGRI detector. A conservative upper limit on the exposure time uncertainty
is $< 2 \%$.  
This leads to uncertainty in the final log N - log S primarily manifest
in the normalization and it should not affect the slope significantly.
Furthermore, the uncertainty in the detection limit will affect mainly the low flux 
end of the Log(N) -- Log(S) distribution. The high flux end is less sensitive
to scatter, since it is based on a larger sky area 
(Fig.~\ref{fig:fluxlimits}). To make a more quantitative assessment, we have
recomputed the ML Log(N) -- Log(S) calculation for scenarios in which the
exposure time -- flux limit curve shifted in amplitude and pivoted about the 700 ks point where we have the highest density of measurements. 
For those scenarios, we found that the inferred
Log(N) -- Log(S) slope varied by less than about 5\%, which is contained 
within the range of our quoted 1-sigma uncertainty. The amplitude varied 
by as much as 7\% in the extreme case, but for the pivoted
cases, by only a few percent. We thus conclude that the maximum uncertainty 
resulting from possible systematics in our effective area correction
is bounded by about 5\% in slope and 7\% in amplitude.

\section{The Local Luminosity Function of AGNs at 20 -- 40 keV}

\begin{figure}
\plotone{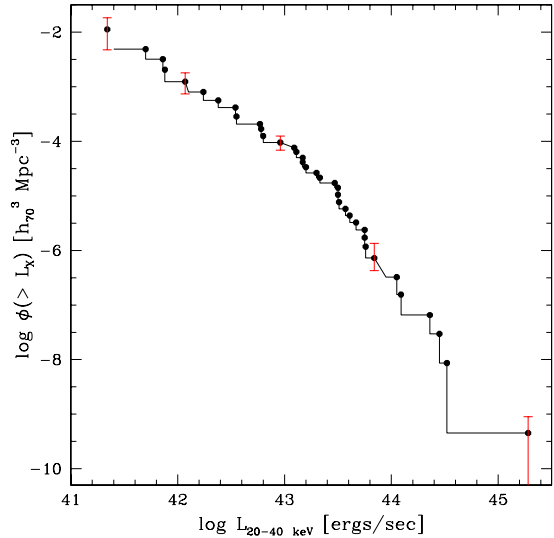}
\caption[]{Cumulative luminosity function of {\it INTEGRAL} AGNs with a detection significance $> 5 \sigma$. Blazars have been excluded. The density $\phi$ describes the number of objects per $\rm Mpc^3$ above a given luminosity $L_X$.
As an example some error bars indicating the Poissonian error are shown. Note that the errors are not independent of each other.
}

\label{fig:lumicum}
\end{figure}

\begin{figure}
\plotone{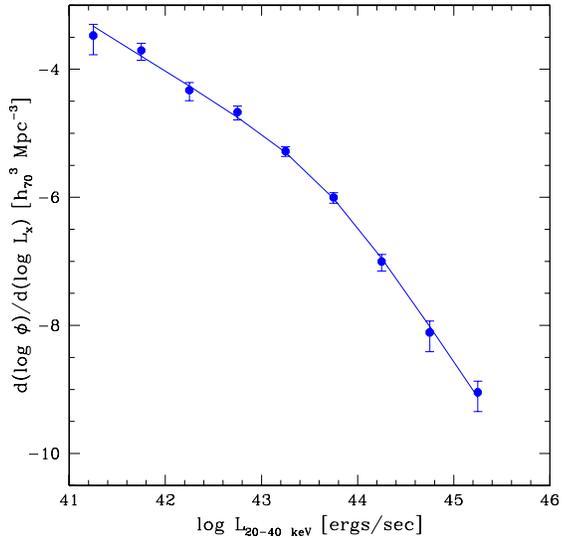}
\caption[]{Differential luminosity function of AGNs with $\Delta \log L_X = 0.5$ binning. The line shows a fit to a smoothly connected two power-law function with a turnover luminosity at $L_* = 2.4 \times 10^{43} \rm \, ergs \, s^{-1}$.}

\label{fig:lumidiff}
\end{figure}

The complete sample of {\it INTEGRAL} AGNs with a detection significance $\ge 5 \sigma$ also allows us to derive the density of these objects in the local Universe as a function of their luminosity. In order to derive the density of objects above a given luminosity, one has to determine for each source in a complete sample the space volume in which this source could have been found considering both the flux limit of each survey field and the flux of the object. 
We have again used the correlation between exposure time and flux limit as discussed in the previous section in order to assign a $5\sigma$ flux limit to each survey field. Then the maximum redshift $z_{max}$ at which an object with luminosity $L_X$ would have been detectable in each sky element was used to compute the total accessible volume 
\begin{equation}
V_a = \sum\limits_{i=1}^{N} \frac{\Omega_i}{4\pi} V_i[z_{max,i}(L_X)]
\end{equation}
with $N$ being the number of sky elements in which the object would have been detectable and $\Omega_i$ the solid angle covered by sky element $i$, and $V_i$ the enclosed volume based on the maximum redshift at which the object could have been detected in this sky element. 
Figure~\ref{fig:lumicum} shows the cumulative luminosity function for 38 {\it INTEGRAL} detected ($\ge 5\sigma$) AGNs in the 20 -- 40 keV energy band. Here the density $\phi$ describes the number of objects per $\rm Mpc^3$ above a given luminosity $L_X$: $\phi = \sum\limits_{i=1}^{K} V_{a,i}^{-1}$ with $K$ being the number of objects with luminosities $> L_X$.  
Blazars have been excluded because their emission is not isotropic. The redshifts in the sample range from $z = 0.001$ to $z = 0.129$ with an average redshift of $\bar{z} = 0.022$. Thus the luminosity function is truly a local one. 
Figure~\ref{fig:lumidiff} shows the luminosity function in differential form. In this presentation the data points are independent of each other. In case one of the luminosity bins would suffer from incompleteness compared to the other bins, this would result in a break or dip in the differential luminosity function. The errors are based on the number of objects contributing to each value. The differential XLF also shows, like the cumulative one, a turnover around $L_X = (5 - 10) \times 10^{43} \rm \, ergs \, s^{-1}$.

Because our study 
is based solely on low redshift objects, we are not able to constrain models involving evolution with redshift. Nevertheless we can compare the XLF presented here with model predictions from previous investigations. XLFs are often fit by a smoothly connected two power-law function of the form \cite{SXLF}

\begin{equation}
\frac{d \phi(L_X, z=0)}{d \log L_X} = A \, {\left[ {\left( \frac{L_X}{L_*} \right)}^{\gamma_1} + {\left( \frac{L_X}{L_*}\right)}^{\gamma_2} \right]}^{-1}
\label{lumifunc}
\end{equation}

We fit this function using a least-squares method applying the Levenberg-Marquardt algorithm (Marquardt 1963). The best fit values we obtained are 
$A = 0.7 {+1.5 \atop -0.5} \times 10^{-5} \, h_{70}^3 \, \rm Mpc^{-3}$, $\gamma_1 = 0.80 \pm 0.15$, $\gamma_2 = 2.11 \pm 0.22$, and $\log L_* = 43.38 \pm 0.35$ with $L_*$ in units of $h_{70}^{-2} \rm \, ergs \, s^{-1}$.
The $1\sigma$ errors have been determined by applying a Monte Carlo simulation which simultaneously takes into account the flux errors on the individual sources, the error induced by deriving an average luminosity per bin, and the statistical error of the density based on the number of objects contributing to the density value. Each simulated data set included 9 luminosity values with a density value for each of them. These values where then fit by the smoothly connected two power-law function as described above. The scatter in the resulting parameters gave the error estimates as shown above.
  
The parameter values describing the differential luminosity function are consistent with values derived from the $2 - 10 \rm \, keV$ XLF of AGNs as shown by e.g. Ueda et al. (2003), Franca et al. (2005), and Shinozaki et al. (2006). For example the work by Ueda et al. (2003) reveals for a pure density evolution model the same values (within the error bars) for $\gamma_1$ and $\gamma_2$, but a higher $\log L_* = 44.11 \pm 0.23$. The higher value  can be easily explained by the different energy bands applied. A single power law with photon index of $\Gamma = 2$ in the range $2 - 40 \rm \, keV$ would lead to $L_{(2 - 10 \rm \, keV)} / L_{(20 - 40 \rm \, keV)} = 2.3$, assuming no intrinsic absorption. 
This has, of course, no implications for the XLF at higher redshifts.
The values are also consistent with the luminosity function for AGNs in the $3 - 20 \rm \, keV$ band as derived by Sazonov \& Revnivtsev (2004) from the {\it RXTE} all-sky survey.
 
\begin{figure}
\plotone{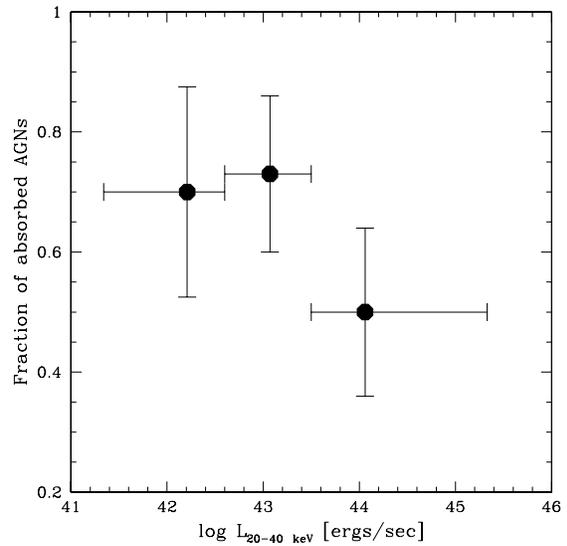}
\caption[]{Fraction of absorbed AGNs ($N_H \ge 10^{22} \rm \, cm^{-2}$) versus luminosity. The position of the data point along the luminosity axis indicates the average luminosity in this bin, while the error bars in luminosity indicate the range of luminosities covered.}

\label{fig:fracabs}
\end{figure}

Information about intrinsic absorption is available for 32 of the 38 objects (89\%) from soft X-ray observations. This enables us to derive the luminosity function for absorbed ($N_H \ge 10^{22} \rm \, cm^{-2}$) and unabsorbed sources, as shown in Figure~\ref{fig:lumiabsorbed}. The absorbed sources have a higher density than the unabsorbed sources at low luminosities, while this trend is inverted at high luminosities. The luminosity where both AGN types have similar densities is about $L_{(20 - 40 \rm \, keV)} = 3 \times 10^{43} \rm \, erg \, s^{-1}$. This tendency is also evident when comparing the fraction of absorbed AGNs with the luminosity in the three luminosity bins depicted in Figure~\ref{fig:fracabs}. The luminosity bins have been chosen so that an equal number of objects are contained in each bin. The position of the data point along the luminosity axis indicates the average luminosity in this bin, while the error bars in luminosity indicate the range of luminosities covered. A comparable trend has been seen also below 10 keV. In a recent study of {\it HEAO-1} data of 49 AGNs, Shinozaki et al. (2006) showed that the XLF for absorbed AGNs drops more rapidly ($\gamma_2 = 3.34 {+0.90 \atop -0.65}$) at higher luminosities than that of unabsorbed AGNs ($\gamma_2 = 2.34 {+0.24 \atop -0.22}$).

Based on the luminosity function, the contribution of the AGNs to the total X-ray emissivity $W$ can be estimated \cite{RXTENGC}. This can be done by simply multiplying the XLF by the luminosity in each bin and integrating over the range of luminosities ($10^{41} {\rm \, ergs \, s^{-1}} < L_{20 - 40 \rm \, keV} < 10^{45.5} \rm \, ergs \, s^{-1}$).
This results in $W_ {20 - 40 \rm \, keV}(> 10^{41} \rm \, ergs \, s^{-1}) = (2.8 \pm 0.8) \times 10^{38} \rm \, ergs \, s^{-1} \, h_{70}^3 \, Mpc^{-3}$.  Please note that absorption does not affect the luminosities in this energy range and therefore the values given here are intrinsic emissivities.

\section{Discussion}

A simple power-law model fitted to the number flux distribution (Fig.~\ref{fig:logNlogS5}) has a slope of $\alpha = 1.66 \pm 0.11$. 
Even though the difference from the Eucledian value is not statistically significant, at a $1.5 \sigma$ level, a deviation from this value could have two reasons. The difference might indicate that the area density at the low flux end of the distribution has been slightly overcorrected. One has to keep in mind that only a few sources derived from a small area of the sky are constraining the low flux end. Another reason for the difference could be that the distribution of AGNs in the very local universe is not isotropic, caused e.g. by the local group and other clustering of galaxies.
Krivonos et al. (2005) studied the extragalactic source counts as observed by {\it INTEGRAL} in the 20-50 keV energy band in the Coma region. 
Based on 12 source detections they determine a surface density of $(1.4 \pm 0.5) \times 10^{-2} \rm \, deg^{-2}$ above a threshold of $10^{-11} \rm \, ergs \, cm^{-2} \, s^{-1}$ in the $20-50 \rm \, keV$ energy band, where we get a consistent value of $(1.2 \pm 0.2) \times 10^{-2} \rm \, deg^{-2}$.
Comparing the total flux of all the objects in the AGN sample ($f_{20 - 40 \rm \, keV} = 2.6 \times 10^{-10} \rm \, ergs \, cm^{-2} \, s^{-1} \, sr^{-1}$) with the flux of the X-ray background as presented by Gruber et al. (1999) shows that the {\it INTEGRAL} AGN account only for about 1\% of the expected value. This is expected when taking into account the high flux limit of our sample: La Franca et al. (2005) have shown that objects with $f_{2 - 10 \rm \, keV} > 10^{-11} \rm \, ergs \, cm^{-2} \, s^{-1}$ contribute less than 1\% to the CXB in the $2 - 10 \rm \, keV$ energy range. This flux limit extrapolates to the faintest flux in our sample of $f_{20 - 40 \rm \, keV} = 5.6 \times 10^{-12} \rm \, ergs \, cm^{-2} \, s^{-1}$ for a $\Gamma = 1.9$ power law spectrum.

We compared the unabsorbed emissivity per unit volume of our objects $W_ {20 - 40 \rm \, keV}(> 10^{41} \rm \, ergs \, s^{-1}) = 2.8 \times 10^{38} \rm \, ergs \, s^{-1} \, h_{70}^3 \, Mpc^{-3}$ with that observed by {\it RXTE} in the 3--20 keV band.  Assuming an average power law of $\Gamma = 2$, the extrapolated value is $W_ {3 - 20 \rm \, keV}(> 10^{41} \rm \, ergs \, s^{-1}) = (7.7 \pm 2.2) \times 10^{38} \rm \, ergs \, s^{-1} \, h_{70}^3 \, Mpc^{-3}$, which is a factor of 2 larger than the value measured by {\it RXTE} \cite{RXTENGC} but consistent within the $1\sigma$ error. If we apply the conversion to the $2 - 10 \rm \, keV$ energy band, 
we derive the intrinsic emissivity $W_ {2 - 10 \rm \, keV} = (6.4 \pm 1.8) \times 10^{38} \rm \, ergs \, s^{-1} \, Mpc^{-3}$, consistent with the value derived from the {\it HEAO-1} measurements 
($W = (5.9 \pm 1.2) \times 10^{38} \rm \, ergs \, s^{-1} \, Mpc^{-3}$; Shinozaki et al. 2006). This showed that the local X-ray volume emissivity in the 2--10 keV band is consistent with the emissivity from AGNs alone. 
It has to be pointed out that the value derived from our sample and the one based on {\it HEAO-1} data are higher than the one based on the {\it RXTE} All-Sky Survey  ($W = (2.7 \pm 0.7) \times 10^{38} \rm \, ergs \, s^{-1} \, Mpc^{-3}$; Sazonov \& Revnivtsev 2004). 

\begin{figure}
\plotone{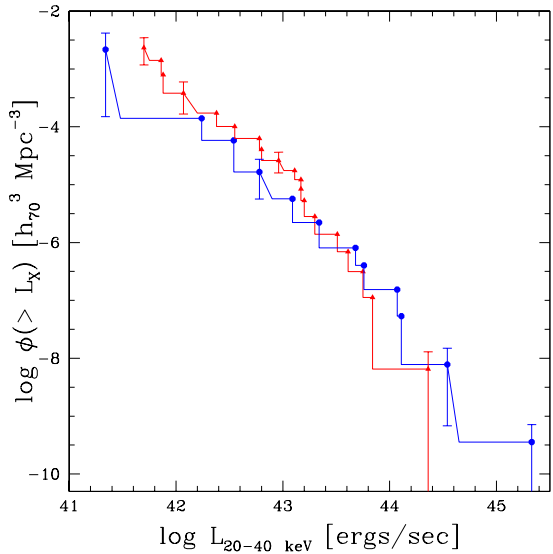}
\caption[]{Cumulative AGN luminosity function for 19 absorbed ($N_H \ge 10^{22} \rm \, cm^{-2}$; triangles) and 12 unabsorbed sources (octagons). As an example some error bars indicating the Poissonian error are shown.
}

\label{fig:lumiabsorbed}
\end{figure}

The luminosity function derived from the {\it INTEGRAL} $20 - 40 \rm \, keV$ AGN sample appears to be consistent with the XLF in the $2 - 20 \rm \, keV$ range. A turnover in the XLF at $\simeq 2.4 \times 10^{43} \rm \, ergs \, s^{-1}$ is observed (Fig.~\ref{fig:lumidiff}). Below this luminosity also the fraction of absorbed AGNs starts to exceed that of the unabsorbed ones, although the effect is significant at only on a $1\sigma$ level (Fig.~\ref{fig:lumiabsorbed}). Both effects have been seen also in the $2 - 10 \rm \, keV$ \citep{Ueda03,LaFranca,HEAO-1XLF} and in the $3 - 20 \rm \, keV$ band \cite{RXTENGC}. This implies that we do detect a similar source population as at lower energies.

If a larger fraction of absorbed AGNs is necessary to explain the cosmic X-ray background at $\sim 30 \rm \, keV$ as indicated by {\it HEAO 1} A-4 measurements \cite{Gruber}, the fraction of absorbed sources could be correlated with redshift. 
It has for example been proposed that there is an evolution of the population leading to a higher fraction of absorbed sources at higher redshifts. It should be noted however that this effect is not clearly detectable in the $2 - 10 \rm \, keV$ range. The fraction of absorbed sources seems to depend on luminosity \citep{Ueda03,AGNunification}, as is also seen in the $20 - 40 \rm \, keV$ band (Fig.~\ref{fig:fracabs}). But some studies come to the conclusion that there is no evolution of intrinsic $N_H$ \citep{Ueda03,AGNunification}, while others find the fraction of absorbed sources increasing with redshift \cite{LaFranca}. La Franca et al. also find that a combination of effects (the fraction of absorbed AGNs decreases with the intrinsic X-ray luminosity, and increases with the redshift) can be explained by a luminosity-dependent density evolution model. They further show that the luminosity function of AGNs with low luminosities as those presented here peaks at $z \sim 0.7$ while high luminosity AGNs peak at $z \sim 2$. Unified models also predict, depending on the applied model, a fraction of absorbed AGNs of $0.6 - 0.7$ compared to the total population for high-flux low-redshift objects \cite{AGNunification}. 
Worsley et al. (2005) examined {\it Chandra} and {\it XMM-Newton} deep fields and come to the conclusion that the missing CXB component is formed by highly obscured AGNs at redshifts $\sim 0.5 -1.5$ with column densities of the order of $f_X = 10^{23} - 10^{24} \rm \, cm^{-2}$. 
Evidence for this scenario is also found in a study of {\it Chandra} and {\it Spitzer} data \cite{SWIRE}. Combining multiwavelength data, this work estimates a surface density of $25 \rm \, AGN \, deg^{-2}$ in the infrared in the $0.6 \rm \, deg^2$ {\it Chandra}/SWIRE field, and only $33 \%$ of them are detected in the X-rays down to $f_{0.3-8 \rm \, keV} = 10^{-15} \rm \, ergs \, cm^{-2} \, s^{-1}$. The work also indicates a higher abundance of luminous and Compton-thick AGNs at higher redshifts ($z \gg 0.5$).
This source population would be missed by the study presented here, because of the low redshifts (${\bar z} = 0.022$) of the {\it INTEGRAL} AGNs.

Several studies \citep{Ueda03,AGNunification} propose that the absorbed AGNs needed to explain the CXB should be Compton thick, and therefore would have been missed at $2 - 10 \rm \, keV$. This argument does not hold for the {\it INTEGRAL} observations, where the impact of absorption is much less severe than at lower energies. The effect on the measured flux of a source with photon index $\Gamma = 2$ for Compton thick absorption ($N_H = 10^{24} \rm \, cm^{-2}$) is only a 5\% decrease in flux (40\% for $N_H = 10^{25} \rm \, cm^{-2}$). It is therefore unlikely that many Compton-thick objects have been missed by the {\it INTEGRAL} studies performed to date. One possibility would be, that they are among the newly discovered sources found by {\it INTEGRAL}. The fraction of unidentified objects among the {\it INTEGRAL} discovered sources is approximately $50 \%$. Eight such sources without cross-identification have a significance above $5\sigma$ in the data set discussed here. Thus, if they are ultimately identified as AGNs, they would have to be considered in this study. It should be pointed out though, that most of the sources discovered by {\it INTEGRAL} are located close to the Galactic plane and are more likely to belong to the Galaxy: the Second IBIS/ISGRI Soft Gamma-Ray Survey Catalog \cite{Bird06} lists 55 new sources detected by {\it INTEGRAL}, of which 93\% are located within $-10^\circ < b < +10^\circ$. Among these 55 sources, 3 are listed as extragalactic sources, 18 are of Galactic origin, and 29 have not been identified yet. 

\begin{figure}
\plotone{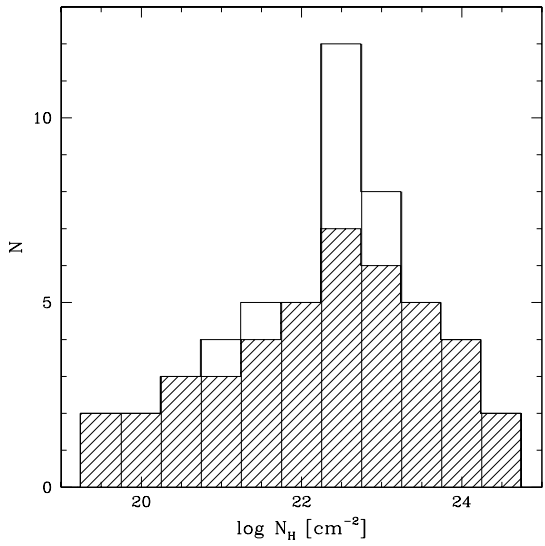}
\caption[]{Distribution of intrinsic absorption for all {\it INTEGRAL} AGNs (blazars excluded), as measured in the soft X-rays. The shaded area shows the reliable measurements, the other values are based on comparison of {\it ROSAT} All-Sky Survey and {\it INTEGRAL} data. Lower limits on absorption have been excluded.}

\label{fig:nhhisto}
\end{figure}

In addition, those objects which have been classified as AGNs based on soft X-ray and/or optical follow-up studies, are no more likely to be Compton-thick objects than the overall AGN population studied here. Only four AGNs (NGC 1068, NGC 4945, MRK 3, Circinus galaxy) detected by {\it INTEGRAL} have been proven to be Compton thick objects so far, and none of them showed absorbtion of $N_H > 5 \times 10^{24} \rm \, cm^{-2}$. In order to clarify this point, observations at soft X-rays of those objects without information about intrinsic absorption are required for all {\it INTEGRAL} detected AGNs (Tab.~\ref{catalog}). At present 23 \% of the {\it INTEGRAL} AGN are missing absorption information. 
A first indication of what the absorption in these sources might be, can be derived from comparison of the {\it INTEGRAL} fluxes with {\it ROSAT} All-Sky Survey (RASS) Faint Source Catalogue data \cite{RASSFSC}. In order to do so we assumed a simple power law with photon index $\Gamma = 2.0$ between the {\it ROSAT} $0.1 - 2.4 \rm \, keV$ band and the {\it INTEGRAL} $20 - 40 \rm \, keV$ range and fit the absorption. In the six cases where no detection was achieved in the RASS, an upper limit of $f_{(0.1 - 2.4 \rm \, keV)} \le 10^{-13} \rm \, ergs \, cm^{-2} \, s^{-1}$ has been assumed, resulting in a lower limit for the absorption $N_H > (5 - 11) \times 10^{22} \rm \, cm^{-2}$. In Fig.~\ref{fig:nhhisto} we show the distribution of intrinsic absorption. It has to be pointed out that the estimated values can only give an idea about the distribution of intrinsic absorption and should not be taken literally, as the spectral slope between the measurements is unknown and the observations are not simultaneous. Nevertheless apparently none of the RASS detections and non-detections requires an intrinsic absorption of $N_H > 2\times 10^{23} \rm \, cm^{-2}$. Therefore it appears unlikely that a significant fraction of {\it INTEGRAL} AGNs will show an intrinsic absorption $N_H > 10^{24} \rm \, cm^{-2}$. 
However, if we assume that the RASS non-detections are all Compton thick AGNs, the fraction of this class of sources rises from 6\% to 16\% when considering all 63 non-blazar AGNs seen by {\it INTEGRAL}, and from 8\% to 13\% for the complete sample with 38 objects. This is in good agreement with the fraction of 11\% of Compton thick AGN as seen in the {\it Swift}/BAT survey \cite{BATsurvey}.
The picture is less clear when referring to the optical classification. Here the {\it INTEGRAL} survey finds 12 Seyfert 1 (33\%), 14 Seyfert 2, and 10 intermediate Seyfert 1.5 in the complete sample, while the {\it Swift}/BAT survey contains only 20\% of type 1 Seyfert galaxies. It should be pointed out though that the classification based on intrinsic absorption gives a more objective criterion in order to define AGN subclasses than the optical classification with its many subtypes. 
The finding of the BAT survey that virtually all sources with $\log L_X < 43.5$ are absorbed, cannot be confirmed by our study, in which we detect a fraction of $33 \%$ of sources with $N_H < 10^{22} \rm \, cm^{-2}$ among the sources below this luminosity. This also reflected in the observation that although the absorbed sources become more dominant below this luminosity, the trend is not overwhelmingly strong (Fig.~\ref{fig:lumiabsorbed}).



Most investigations to date have been focused
on the X-rays below 20 keV, and {\it INTEGRAL} can add
substantial information to the nature of bright AGNs in the local
Universe. Considering the expected composition of the hard X-ray background, it does not currently appear that the population detected by {\it INTEGRAL} can explain the peak at 30 keV, as Compton thick AGNs are apparently less abundant than expected \cite{AGNunification}. But this picture might change if we assume that  all {\it INTEGRAL} AGNs lacking soft X-ray data and without counter parts in the RASS to be Compton thick.  
In addition the sample presented here might be still too small to constrain the
fraction of obscured sources, and the missing Compton thick AGNs could be detectable when studying sources with $f_{(20 - 40 \rm \, keV)} < 10^{-11} \rm \, ergs \, cm^{-2} \, s^{-1}$.

\section{Conclusions}

The extragalactic sample derived from the {\it INTEGRAL} public data archive
comprises 63 low redshift Seyfert galaxies ($\langle z \rangle = 0.022 \pm 0.003$) and 8~blazars in the hard X-ray domain. 
Two galaxy clusters are also detected, but no star-burst galaxy has been as yet. This {\it INTEGRAL} AGN sample is thus the largest one presented so far. 38 of the Seyfert galaxies form a complete sample with significance limit of $5 \sigma$. 

The number flux distribution is approximated by a power-law with a slope of $\alpha = 1.66 \pm 0.11$. Because of the high flux limit of our sample the objects account in total for less than $1\%$ of the $20 - 40 \rm \, keV$ cosmic X-ray background. The emissivity of all AGNs per unit volume $W_ {20 - 40 \rm \, keV}(> 10^{41} \rm \, ergs \, s^{-1}) = 2.8 \times 10^{38} \rm \, ergs \, s^{-1} \, h_{70}^3 \, Mpc^{-3}$ appears to be consistent with the background estimates in the 2--10 keV energy band based on the cross-correlation of the {\it HEAO 1}-A2 map with {\it IRAS} galaxies \cite{Miyaji94}.

The luminosity function in the $20 - 40 \rm \, keV$ energy range is consistent with that measured in the $2 - 20 \rm \, keV$ band. Below the turnover luminosity of $L_* = 2.4 \times 10^{43} \rm \, ergs \, s^{-1}$ the absorbed AGNs become dominant over the unabsorbed ones. 
The fraction of Compton thick AGNs with known intrinsic absorption is found to be small ($8\%$) in our AGN sample.
For the sources without reliable absorption information we derived an estimate from the comparison with {\it ROSAT} All-Sky Survey data and find that the data do not require additional Compton thick objects within the sample presented here. 
It has to be pointed out though, that the sources without RASS counterpart could be Compton thick which would increase the ratio of this source type to 13\% in the complete sample.
Evolution of the source population can play a major role in the sense that the fraction of absorbed sources among AGNs might be correlated with redshift, as proposed for example by Worsley et al. (2005). 

Over the life time of the {\it INTEGRAL} mission we expect to detect of the order of 200 AGNs. Combining these data with the studies based on {\it Swift}/BAT, operating in a similar energy band as IBIS/ISGRI, will further constrain the hard X-ray luminosity function of AGNs. But we will still be limited to the relatively high flux end of the distribution. Because of this {\it INTEGRAL} and {\it Swift}/BAT will most likely not be able to test evolutionary scenarios of AGNs and thus will be inadequate to explain the cosmic X-ray background at $E > 20 \rm \, keV$. Future missions with larger collecting areas and/or focusing optics 
will be required to answer the question of what dominates the Universe in the hard X-rays.

\begin{acknowledgements}

VB would like to thank Olaf Wucknitz for providing software to handle the $\Lambda_0 > 0$ cosmology.
This research has made use of the NASA/IPAC Extragalactic Database (NED) which is operated by the Jet Propulsion Laboratory, of data obtained from the High Energy Astrophysics Science Archive Research Center (HEASARC), provided by NASA's Goddard Space Flight Center, and of the SIMBAD Astronomical Database which is operated by the Centre de Donn\'ees astronomiques de Strasbourg. This research has also made use of the Tartarus (Version 3.1) database, created by Paul O'Neill and Kirpal Nandra at Imperial College London, and Jane Turner at NASA/GSFC. Tartarus is supported by funding from PPARC, and NASA grants NAG5-7385 and NAG5-7067.
\end{acknowledgements}

%
%
%

%
%
\begin{deluxetable}{lclrrrrcccc}
\tabletypesize{\scriptsize}
\tablecaption{{\it INTEGRAL} AGN catalog\label{catalog}}
\tablehead{
\colhead{Name} & \colhead{Type} & \colhead{z} & \colhead{R.A.} & \colhead{Decl.} &
\colhead{exp.\tablenotemark{a}} & \colhead{ISGRI} &
\colhead{$f_X\tablenotemark{b}$} &
\colhead{$\log L_{20-40 \rm \, keV}$} &
\colhead{$N_H\tablenotemark{c,f}$} &
\colhead{Ref.\tablenotemark{f}}\\
\colhead{} & \colhead{} & \colhead{} & \colhead{(J2000.0)} & \colhead{(J2000.0)} & \colhead{(ks)} & \colhead{$(\sigma)$} & \colhead{} & \colhead{($\rm ergs \; s^{-1})$} & \colhead{} & \colhead{}\\
}
\startdata
1ES 0033+595    & BL Lac & 0.086    & 00 35 53 &  +59 50 05 & 1449 &  3.5 & 0.37 & 43.83  & 0.36\tablenotemark{1} & 2\\ 
NGC 788         & Sy 1/2 & 0.0136   & 02 01 06 & --06 48 56 &  311 & 10.7 & 2.98 & 43.09  & $<0.02\tablenotemark{3}$ & 4 \\ 
IGR J02097+5222 & Sy 1 & 0.0492 &    02 09 46 &  +52 22 48 &    26 &   4.9 & 3.8 &  44.34 & ? & 30\\
NGC 1068        & Sy 2   & 0.003793 & 02 42 41 & --00 00 48 &  311 &  4.3 & 0.93 & 41.47 & $>150\tablenotemark{5}$ & 4 \\ 
QSO B0241+62    & Sy 1   & 0.044557 & 02 44 58 &  +62 28 07 &   43 &  3.4 & 2.02 & 43.97 & 1.5\tablenotemark{6} & 7\\ 
NGC 1142        & Sy 2   & 0.028847 & 02 55 12 & --00 11 01 &  311 &  5.5 & 1.58 & 43.48 &  ? & 8\\ 
NGC 1275        & Sy 2   & 0.017559 & 03 19 48 &  +41 30 42 &  264 &  8.4 & 1.89 & 43.12 & 3.75\tablenotemark{4} & 4\\ 
3C 111          & Sy 1   & 0.048500 & 04 18 21 &  +38 01 36 &   67 & 10.0 & 6.27 & 44.54 & 0.63\tablenotemark{3} & 4 \\ 
UGC 3142        & Sy 1   & 0.021655 & 04 43 47 &  +28 58 19 &  247 & 16.8 & 5.46 & 43.76 & ? & 2\\ 
LEDA 168563     & Sy 1   & 0.0290   & 04 52 05 &  +49 32 45 &   28 &  2.8 & 2.27 & 43.64 & ? & 2\\ 
MCG +8--11--11  & Sy 1.5 & 0.020484 & 05 54 54 &  +46 26 22 &   21 &  6.2 & 6.07 & 43.76 & $< 0.02\tablenotemark{6}$ & 4\\ 
MRK 3           & Sy 2   & 0.013509 & 06 15 36 &  +71 02 15 &  472 & 15.9 & 3.65 & 43.17 & 110\tablenotemark{6} & 4\\ 
MRK 6           & Sy 1.5 & 0.018813 & 06 52 12 &  +74 25 37 &  482 &  8.7 & 2.01 & 43.21 & 10\tablenotemark{6} & 4\\ 
S5 0716+714     & BL Lac & 0.3\tablenotemark{d}  & 07 21 53 &  +71 20 36 &  482 & 0.7 & 0.14 & 44.41\tablenotemark{d} & $<0.01\tablenotemark{3}$ & 9 \\ 
ESO 209--12     & Sy 1.5 & 0.040495 & 08 01 58 & --49 46 36 & 1543 &  6.7 & 0.86 & 43.52 & ? & 7 \\ 
FRL 1146        & Sy 1   & 0.031789 & 08 38 31 & --35 59 35 &  849 &  3.6 & 0.60 & 43.15 & ? & 7 \\ 
S5 0836+710     & FSRQ   & 2.172    & 08 41 24 &  +70 53 42 &  391 &  6.4 & 1.73 & 47.79 & 0.11\tablenotemark{3} & 9 \\ 
MCG--05--23--16 & Sy1.9  & 0.008486 & 09 47 40 & --30 56 56 &    2 &  2.3 & 11.20 &43.25 & 1.6\tablenotemark{16} & 16\\ 
IGR J10404--4625& Sy 2   & 0.0237   & 10 40 22 & --46 25 26 &   46 &  1.5 & 0.67 & 42.93 & ? & 10\\ 
NGC 3783 & Sy 1 & 0.00973 & 11 39 02 & --37 44 19 & 18\tablenotemark{e} & 5.6 & 6.2 & 43.11 & $<0.4\tablenotemark{3}$ & 23\\
IGR J12026--5349& AGN    & 0.028    & 12 02 48 & --53 50 08 &  191 &  5.5 & 1.86 & 43.52 & 2.2\tablenotemark{11} & 11\\ 
NGC 4051        & Sy 1.5 & 0.002336 & 12 03 10 &  +44 31 53 &  443 &  8.4 & 1.80 & 41.34 & $<0.01\tablenotemark{6}$ & 4\\ 
NGC 4151        & Sy 1.5 & 0.003320 & 12 10 33 &  +39 24 21 &  483 &163.3 &26.13 & 42.80 & 6.9\tablenotemark{12} & 12\\ 
NGC 4253        & Sy 1.5 & 0.012929 & 12 18 27 &  +29 48 46 &  715 &  6.1 & 0.93 & 42.54 & 0.8\tablenotemark{6} & 4\\ 
4C +04.42       & BL Lac & 0.965    & 12 22 23 &  +04 13 16 &  690 &  4.5 & 0.80 & 46.58 & ? & 7\\ 
NGC 4388        & Sy 2   & 0.008419 & 12 25 47 &  +12 39 44 &  215 & 34.8 & 9.54 & 43.18 & 27\tablenotemark{13} & 13\\ 
NGC 4395        & Sy 1.8 & 0.001064 & 12 25 49 &  +33 32 48 &  739 &  5.1 & 0.56 & 40.14 & 0.15\tablenotemark{3} & 4\\ 
3C 273          & Blazar & 0.15834  & 12 29 07 &  +02 03 09 &  655 & 34.2 & 5.50 & 45.58 & 0.5\tablenotemark{4} & 14\\ 
NGC 4507        & Sy 2   & 0.011801 & 12 35 37 & --39 54 33 &  152 & 14.9 & 6.46 & 43.30 & 29\tablenotemark{6} & 4 \\ 
IGR J12391--1612& Sy 2   & 0.036    & 12 39 06 & --16 10 47 &   83 &  1.4 & 3.46 & 44.02 & 1.9\tablenotemark{11} & 11\\ 
NGC 4593        & Sy 1   & 0.009000 & 12 39 39 & --05 20 39 &  723 & 20.1 & 3.31 & 42.78 & 0.02\tablenotemark{6} & 4\\ 
IGR J12415--5750& Sy 2   & 0.024    & 12 41 24 & --57 50 24 &  440 &  1.1 & 0.33 & 42.64 & ? & 15 \\ 
3C 279          & Blazar & 0.53620  & 12 56 11 & --05 47 22 &  497 &  3.6 & 0.82 & 45.97 & $\le 0.13\tablenotemark{3}$ & 4\\ 
Coma Cluster    & GClstr & 0.023100 & 12 59 48 &  +27 58 48 &  516 &  7.2 & 1.09 & 43.11 & $<0.01\tablenotemark{4}$ & 4\\ 
NGC 4945        & Sy 2   & 0.001878 & 13 05 27 & --49 28 06 &  276 & 33.8 & 9.85 & 41.88 & 400\tablenotemark{6} & 16 \\ 
ESO 323--G077   & Sy 2   & 0.015014 & 13 06 26 & --40 24 53 &  761 &  6.9 & 1.20 & 42.78 & 55\tablenotemark{17} & 17\\ 
IGR J13091+1137 & AGN    & 0.0251   & 13 09 06 &  +11 38 03 &   48 &  2.0 & 1.06 & 43.18 & 90\tablenotemark{11} & 15 \\ 
NGC 5033        & Sy 1.9 & 0.002919 & 13 13 28 &  +36 35 38 &  377 &  4.6 & 1.06 & 41.30 & 2.9\tablenotemark{6} & 4\\ 
Cen A           & Sy 2   & 0.001830 & 13 25 28 & --43 01 09 &  532 &167.4 &32.28 & 42.38 & 12.5\tablenotemark{4} & 16\\ 
MCG--06--30--015& Sy 1   & 0.007749 & 13 35 54 & --34 17 43 &  567 &  4.9 & 0.73 & 41.99 & 7.7\tablenotemark{6} & 4\\ 
4U 1344--60     & Sy 1.5 & 0.012    & 13 47 25 & --60 38 36 &  603 & 16.6 & 2.83 & 43.02 &    5\tablenotemark{29} & 4\\ 
IC 4329A        & Sy 1.2 & 0.016054 & 13 49 19 & --30 18 36 &  440 & 41.7 & 8.19 & 43.68 & 0.42\tablenotemark{6} & 4\\ 
Circinus gal.   & Sy 2   & 0.001448 & 14 13 10 & --65 20 21 &  589 & 58.9 &10.73 & 41.69 & 360\tablenotemark{6} & 16\\ 
NGC 5506        & Sy 1.9 & 0.006181 & 14 13 15 & --03 12 27 &  101 & 12.6 & 4.21 & 42.55 & 3.4\tablenotemark{6} & 4\\ 
NGC 5548        & Sy 1.5 & 0.017175 & 14 18 00 &  +25 08 12 &  211\tablenotemark{e} &  2.4 & 0.71 & 42.67 & 0.51\tablenotemark{6} & 4\\ 
PG 1416--129    & Sy 1   & 0.129280 & 14 19 04 & --13 10 44 &  117 &  8.3 & 4.86 & 45.33 & 0.09\tablenotemark{4} & 4\\ 
ESO 511--G030   & Sy 1   & 0.022389 & 14 19 22 & --26 38 41 &  145 &  5.1 & 1.93 & 43.34 & $<0.05\tablenotemark{17}$ &  17\\ 
IC 4518         & Sy 2   & 0.015728 & 14 57 43 & --43 07 54 &  338 &  2.3 & 0.49 & 42.44 & ? & 4\\ 
IGR J16119--6036& Sy 1   & 0.016    & 16 11 54 & --60 36 00 &  475 &  1.2 & 0.25 & 42.16 & ? & 2\\ 
IGR J16482--3036& Sy 1   & 0.0313   & 16 48 17 & --30 35 08 &  973 &  4.2 & 0.73 & 43.22 & ? & 10\\ 
NGC 6221        & Sy 1/2 & 0.004977 & 16 52 46 & --59 13 07 &  523 &  5.6 & 1.32 & 41.86 & 1\tablenotemark{18} & 4\\ 
Oph Cluster     & GClstr & 0.028    & 17 12 26 & --23 22 33 & 1763 & 30.8 & 4.10 & 43.90 & ? & 25\\
NGC 6300        & Sy 2   & 0.003699 & 17 17 00 & --62 49 14 &  173 & 10.0 & 3.91 & 42.07 & 22\tablenotemark{19} & 16\\ 
GRS 1734--292   & Sy 1   & 0.021400 & 17 37 24 & --29 10 48 & 3332 & 45.9 & 4.03 & 43.62 & 3.7\tablenotemark{4} & 24\\ 
2E 1739.1--1210 & Sy 1   & 0.037    & 17 41 54 & --12 11 52 &  631 &  5.5 & 1.03 & 43.51 & ? & 7\\ 
IGR J18027--1455& Sy 1   & 0.035000 & 18 02 47 & --14 54 55 &  942 & 12.6 & 2.03 & 43.76 & 19.0\tablenotemark{4} & 20\\ 
PKS 1830--211   & Blazar & 2.507    & 18 33 40 & --21 03 40 & 1069 & 12.7 & 2.07 & 48.02 & $\le 0.7\tablenotemark{3}$ & 21 \\ 
ESO 103--G35    & Sy 2   & 0.013286 & 18 38 20 & --65 25 39 &   36 &  4.2 & 2.97 & 43.07 & 19\tablenotemark{28} & 16\\ 
3C 390.3  & Sy 1 & 0.0561 & 18 42 09 & +79 46 17 & 490\tablenotemark{e} & 9.5 & 2.0 & 44.16 & $<0.1$\tablenotemark{27} & 23\\
2E 1853.7+1534  & Sy 1   & 0.084    & 18 56 00 &  +15 38 13 &  761 &  9.3 & 1.74 & 44.48 & ? &10\\ 
1H 1934--063    & Sy 1   & 0.010587 & 19 37 33 & --06 13 05 &  684 &  2.7 & 0.48 & 42.08 & ? & 4\\ 
NGC 6814        & Sy 1.5 & 0.005214 & 19 42 41 & --10 19 25 &  488 & 12.1 & 2.92 & 42.24 & $<0.05\tablenotemark{3}$ & 4\\ 
IGR J19473+4452 & Sy 2   & 0.0539   & 19 47 19 &  +44 49 42 &  969 &  5.9 & 1.05 & 43.86 & 11\tablenotemark{11} & 11\\ 
Cygnus A        & Sy 2   & 0.056075 & 19 59 28 &  +40 44 02 & 1376 & 21.6 & 3.24 & 44.39 & 20\tablenotemark{22} & 4\\ 
MCG +04--48--002& Sy 2    & 0.014206 & 20 28 35 &  +25 44 00 &  187 &  3.1 & 1.10 & 42.70 & ? & 2\\ 
4C +74.26       & AGN    & 0.104   & 20 42 37 &  +75 08 02 &   72\tablenotemark{e} &  1.9 & 0.93 & 42.35 & 0.21\tablenotemark{3} & 23 \\ 
MRK 509         & Sy 1   & 0.034397 & 20 44 10 & --10 43 25 &   73 &  8.6 & 4.66 & 44.11 & $<0.01\tablenotemark{3}$ & 4\\ 
CJF B2116+818   & Sy 1   & 0.086    & 21 14 01 &  +82 04 48 &  192\tablenotemark{e} & 4.8 & 1.8 & 44.51 & $<0.1$\tablenotemark{1} & 23\\
IGR J21247+5058 & AGN & 0.020 & 21 24 39 &  +50 58 26 &  213 & 11.9 & 4.15 & 43.57 & ? & 20\\ 
NGC 7172 & Sy 2 & 0.00868 & 22 02 02 & --31 52 11 & 401\tablenotemark{e} & 15.1 & 3.3 & 47.74 & 9.0\tablenotemark{28} & 23\\ 
3C 454.3        & Blazar &  0.859   & 22 53 58 &  +16 08 54 &  92\tablenotemark{e} & 2.8 & 3.1 & 46.56 & ? & 26\\
MR 2251--178    & Sy 1   & 0.063980 & 22 54 06 & --17 34 55 &  489 &  7.0 & 1.20 & 44.07 & $\le 0.19\tablenotemark{3}$ & 4\\ 
MCG --02--58--022&Sy 1.5 & 0.046860 & 23 04 44 & --08 41 09 &  489 &  3.9 & 1.20 & 43.79 & $\le 0.08\tablenotemark{3}$ & 4\\ 

\enddata

\tablenotetext{a}{IBIS/ISGRI exposure time}
\tablenotetext{b}{flux $f_{(20 - 40 \rm \, keV)}$ in $10^{-11} \, \rm erg \, cm^{-2} \, s^{-1}$}
\tablenotetext{c}{intrinsic absorption in $10^{22} \rm \, cm^{-2}$}
\tablenotetext{d}{tentative redshift}
\tablenotetext{e}{not covered by survey presented here}
\tablenotetext{f}{REFERENCES.--- (1) Donato, Sambruna, Gliozzi 2005; (2) Bassani et al. 2006; (3) Tartarus database; (4) Beckmann et al. 2006a; (5) Matt et al. 1997; (6) Lutz et al. 2004; (7) Bird et al. 2006; (8) Virani et al. 2005; (9) Pian et al. 2005; (10) Masetti et al. 2005; (11) Sazonov et al. 2005; (12) Beckmann et al. 2005; (13) Beckmann et al. 2004; (14) Courvoisier et al. 2003b; (15) Revnivtsev et al. 2006; (16) Soldi et al. 2005; (17) Sazonov \& Revnivtsev 2004; (18) Levenson, Weaver, \& Heckman 2001; (19) Matsumoto, Nava, Maddox et al. 2004; (20) Masetti et al. 2004; (21) De Rosa et al. 2005; (22) Young et al. 2002; (23) This work; (24) Sazonov et al. 2004; (25) Revnivtsev et al. 2004; (26) Pian et al. 2006; (27) Gliozzi, Sambruna, Eracleous 2003; (28) Akylas, Georgantopoulos, Comastri 2001; (29) Piconcelli et al. 2006; (30) Kuiper et al. 2005
}

\end{deluxetable}
%
%
%

%
%
%

\end{document}